\documentclass[twocolumn,pra,showpacs]{revtex4}

\usepackage{graphicx}
\usepackage{amsmath}

\begin{document}

\title{Low-density, one-dimensional quantum gases in a split trap}

\author{Th.~Busch}
\email{thbusch@phys.ucc.ie} 
\author{G.~Huyet}

\affiliation{Department of Physics, National University of Ireland,
             UCC, Cork, Ireland}

\date{\today}

\begin{abstract}
  We investigate degenerate quantum gases in one dimension trapped in
  a harmonic potential that is split in the centre by a pointlike
  potential. Since the single particle eigenfunctions of such a system
  are known for all strengths of the central potential, the dynamics
  for non-interacting fermionic gases and low-density, strongly
  interacting bosonic gases can be investigated exactly using the
  Fermi-Bose mapping theorem. We calculate the exact many-particle
  ground-state wave-functions for both particle species, investigate
  soliton-like solutions, and compare the bosonic system to the
  well-known physics of Bose gases described by the Gross-Pitaevskii
  equation.  We also address the experimentally important questions of
  creation and detection of such states.
\end{abstract}


\maketitle

\section{Introduction}
\label{sec:Introduction}

Low dimensional quantum gases are fastly progressing from being a
solely successful theoretical concept
\cite{Liniger:63,Lieb:63,Girardeau:60,Girardeau:65,Tonks:36} towards
becoming experimentally available systems in the laboratories
\cite{Prentiss:99,Ketterle:01,Zimmermann:01,Birkl:02,Raithel:02}.
Small scale trapping potentials for low temperature samples allow the
creation of degenerate fermionic or bosonic gases, for which the
dynamics along two spatial directions can be 'frozen out'.  Recent
experimental progress has been the observation of
quasi-one-dimensional Bose-condensates showing spatial phase
fluctuations \cite{Lewenstein:01} and the generation of trapping
frequencies in two-dimensional optical lattices of up to about 120kHz
\cite{Esslinger:01}.

At very low temperatures the interaction between two neutral particles
can often be described using a hard core potential that is dependent
on a single parameter only, the $s$-wave scattering length.  The
Tonks-Girardeau (TG) gas is a model for a one-dimensional bosonic gas
made from particles interacting via such a hard-core potential
\cite{Girardeau:60,Tonks:36} and in the limit of point-like
interactions it can be solved exactly using the Fermi-Bose mapping
theorem \cite{Girardeau:60}. This theorem is one of the
particularities of one-dimensional systems and it states that the
properties of a strongly interacting, hard-core bosonic gas can be
calculated by considering a non-interacting fermionic system. The
first examples for which the TG gas model was solved were free space
and box-like systems with periodic boundary conditions, and recently
Wright and Girardeau have calculated its solution in an external
harmonic potential \cite{Wright:00,Triscari:01}.

In order to find the many-particle solutions to a given geometry for
the TG gas (or for non-interacting fermions) one must know the exact
single particle eigenstates.  However, since the list of exactly
solvable single particle problems in quantum mechanics is limited,
there are even fewer many-particle problems that can be exactly
solved.  Here we present a further solvable system, which comprises of
a gas trapped in a harmonic oscillator potential split in the center
by a point-like potential.  This model can, on the one hand, be seen
as a limit of double well situations \cite{Wilkens:97}, which have
recently been very thoroughly investigated for examining weakly
coupled Bose-Einstein condensates \cite{Shenoy:97}. On the other hand
it is a good approximation to the situation where an impurity is
trapped in the center of a gas.

In Fermi-Bose mapping theory the bosonic wave-function can be
calculated directly from an appropriately chosen fermionic one by
symmetrization. Since the symmetry or antisymmetry of a wave function
does not have an influence on the density distribution, the spatial
density profiles for fermionic and bosonic samples in this limit are
indistinguishable.  Therefore, whenever results concerning density
distributions are presented in this paper, they apply to fermionic as
well as to bosonic samples.

The organization of the paper is the following: we will first briefly
introduce the Fermi-Bose mapping theorem and its application to the TG
gas. Then we will introduce the $\delta$-split harmonic oscillator and
present its exact solutions for arbitrary strength of the splitting
potential. After this we will derive the exact many-particle
ground-state wave-functions for the fermionic and the bosonic case in
the limit of a completely split oscillator and investigate the
dynamical behaviour when the central potential strength is changed. To
compare the results to Bose gases in the Gross-Pitaevskii limit, we
investigate a soliton-like solution.  We suggest different methods for
experimentally realizing a system like this and outline a strategy to
observe it.

\section{Fermi-Bose Mapping Theorem}
\label{sec:FermiBoseMapping}

We consider a gas of $N$ neutral bosonic atoms of mass $m$ at zero
temperature in one dimension. The transverse dynamics is assumed to be
'frozen out' by a very stiff harmonic potential. The density of the
gas shall be low enough to fulfill the requirement that the system is
in the Tonks-Girardeau limit \cite{Olshanii:98,Walraven:00}.  In this
limit the interaction between the particles can be treated as
point-like.  Allowing that the atoms are longitudinally trapped in an
external potential, $V(x)$, the many-particle Hamiltonian can be
written as
\begin{equation}
  \label{eq:Hamiltonian}
   \widehat H=\sum_{i=1}^N
              \left[-\frac{\hbar}{2m}\frac{\partial^2}{\partial x_i^2}
                    +V(x_i)\right]
              +g_1\sum_{i<j}^N\delta(|x_i-x_j|).
\end{equation}
Here the first sum represents the single particle Hamiltonian,
$\widehat H_0$, and the second sum accounts for the inter-particle
interaction. The coupling constant is given by
$g_1=-2\hbar^2/2ma_{\text{1D}}$, with the repulsive, one-dimensional $s$-wave
scattering length given by
$a_{\text{1D}}=(-a_\perp^2/2a_{\text{3D}})[1-Ca_{\text{3D}}/a_\perp]$ \cite{Olshanii:98}.
Here $a_{\text{3D}}$ is the three-dimensional $s$-wave scattering length,
$a_\perp$ the size of the ground state of the transversal harmonic
trapping potential and $C=1.4603\dots$.  In the Tonks-Girardeau limit
the one-dimensional coupling between the particles becomes large and
we will consider $g_1\rightarrow\infty$ in the Hamiltonian
\eqref{eq:Hamiltonian}. In this situation, instead of including the
scattering term into the Hamiltonian, it is more convenient to account
for the strong repulsive interactions using a constraint on the
allowed solutions \cite{Triscari:01}
\begin{equation}
  \label{eq:constraint}
  \psi=0\qquad\text{if}\quad|x_i-x_j|=0,\qquad i\neq j.
\end{equation}
Since this constraint is equivalent to the demand of the Pauli
exclusion principle for a gas of spin-less fermions, one can calculate
the bosonic eigenfunctions, $\psi_B$, of the Hamiltonian of
eq.~\eqref{eq:Hamiltonian} from the fermionic eigenfunctions,
$\psi_F$, of the non-interacting Hamiltonian by proper symmetrisation
\begin{equation}
  \label{eq:Symmetrisation}
  \Psi_B(x_1,\dots,x_N)=|\Psi_F(x_1,\dots,x_N)|.
\end{equation}
This procedure is known as the Fermi-Bose mapping theorem
\cite{Girardeau:60,Girardeau:65}. It can be generalized to time
dependent problems.

To find the many-particle wave-function of the interacting Bose-gas we
therefore need to solve the Slater determinant for non-interacting
fermions. This determinant is composed of the eigenstates of the
corresponding single particle Hamiltonian, $\widehat H_0$.  However,
the number of physical systems where the single particle eigenstates
are available analytically is very small.  One example is the
$\delta$-split harmonic oscillator and we will review its single
particle states in the next section.

\section{Eigenstates of the $\delta$--split harmonic trap}
\label{sec:EigenstatesDeltaSplitTrap}

\begin{figure}[tbp]
  \includegraphics[width=\linewidth]{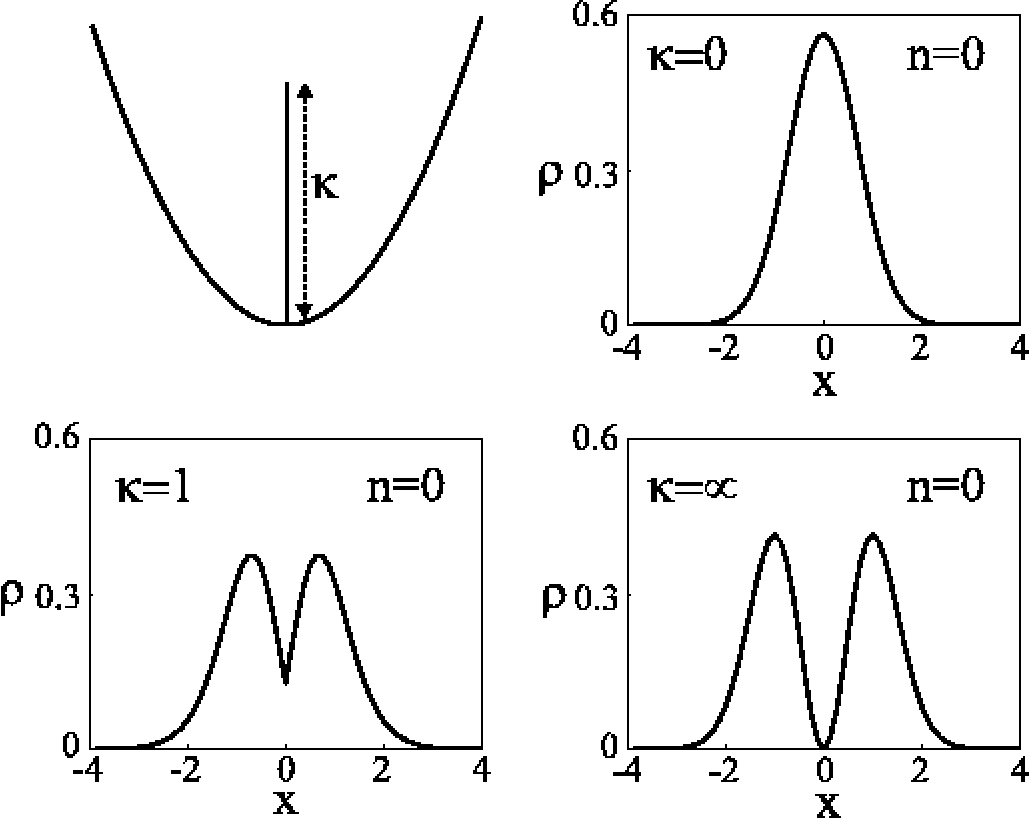}
  \caption{Sketch of the delta split harmonic oscillator and the respective
    single particle ground state ($n=0$) densities for different
    values of $\kappa$. The functions for $\kappa=0, 1, \infty$
    correspond to the energy eigenvalues of
    $E_n=\frac{1}{2}\hbar\omega, 1.08
    \hbar\omega,\frac{3}{2}\hbar\omega$ respectively.}
 \label{fig:GSkappa}
\end{figure}

The single particle Hamiltonian describing a one-dimensional harmonic
trap of frequency $\omega$ with a $\delta$-potential of strength
$\kappa$ in the center is given by
\begin{equation}
  \label{eq:deltaHamiltonian}
  \hat H_0=-\frac{\hbar^2}{2m}\frac{d^2}{d x^2}
           +\frac{1}{2}m\omega^2 x^2+\kappa\delta(x).
\end{equation}
A sketch of the potential is shown in the upper left of
Fig.~\ref{fig:GSkappa}.  One can note immediately that the odd
eigenfunctions, $\psi_{2n+1}$, of the undisturbed harmonic oscillator
($\kappa=0$) are still eigenfunctions of the Hamiltonian $\hat H_0$,
since they are not affected by the potential ($\psi_{2n+1}(0)=0$).  In
particular, they are independent of the strength $\kappa$ of the
$\delta$--function.

To find the even eigenfunctions of the Hamiltonian in
eq.~\eqref{eq:deltaHamiltonian}, let us scale all quantities in units
of the ground state size $a_0=\sqrt{\hbar/2m\omega}$ and the energy
$\epsilon_0=\hbar\omega$ of the harmonic oscillator with $\kappa=0$.
The stationary Schr\"odinger equation then takes the form
\begin{align}
  \label{eq:SE}
  \left(-\frac{d^2}{dx^2}
  +\frac{1}{4}x^2+\tilde\kappa\delta(x)+\epsilon_n\right)\phi_n(x)
  =0,
\end{align}
where we have also introduced the normalized energy
$\epsilon_n=-E_n/(\hbar\omega)$ and the normalized delta function
strength $\tilde\kappa=\kappa a_0/\epsilon_0$. Note the negative sign
of the normalized energy, which has been chosen for easier
mathematical treatment. One can see that for $x>0$ eq.~\eqref{eq:SE}
is equal to the differential equation for the parabolic cylinder
functions and the solutions that have the physically correct
asymptotic behavior are Whittaker's functions, $U(\epsilon_n,x)$,
given by \cite{Abramowitz}
\begin{align}
  \label{eq:Whittaker}
  U(\epsilon_n,&x)=
  \cos\left(\frac{\pi}{4}+\frac{\pi\epsilon_n}{2}\right)Y_1-
  \sin\left(\frac{\pi}{4}+\frac{\pi\epsilon_n}{2}\right)Y_2,\\
  Y_1=&
  \frac{\Gamma\left(\frac{1}{4}-\frac{1}{2}\epsilon_n\right)}
       {\sqrt\pi 2^{(\frac{1}{4}+\frac{1}{2}\epsilon_n)}}
  e^{-\frac{1}{4}x^2}
  M\left(\frac{1}{4}+\frac{1}{2}\epsilon_n,\frac{1}{2},\frac{1}{2}x^2\right),\\
  Y_2=&
  \frac{\Gamma\left(\frac{3}{4}-\frac{1}{2}\epsilon_n\right)}
       {\sqrt\pi 2^{(-\frac{1}{4}+\frac{1}{2}\epsilon_n)}}
  e^{-\frac{1}{4}x^2}x
  M\left(\frac{3}{4}+\frac{1}{2}\epsilon_n,\frac{3}{2},\frac{1}{2}x^2\right).
\end{align}
The $M(a,b,c)$ are the confluent hypergeometric functions. To first
construct the symmetric eigenfunctions of the Hamiltonian in
eq.~\eqref{eq:deltaHamiltonian} on the domain $x\neq 0$ we can now
define \cite{Wilkens:98}
\begin{equation}
  \label{eq:16}
  \phi_n(x)=CU(\epsilon_n,|x|),
\end{equation}
with the normalization constant $C$ chosen such that the
wave--function is normalized to one. Finally, evaluating the
continuity condition at $\phi_n(x=0)$
\begin{equation}
  \label{eq:ContEq}
  \frac{d}{dx}\phi_n(0^+)-\frac{d}{dx}\phi_n(0^-)=\tilde\kappa\phi_n(0),
\end{equation}
we get an implicit relation between the strength of the
$\delta$-function and the eigenvalues of the Hamiltonian
\begin{equation}
  \label{eq:19}
  \frac{\Gamma\left(\frac{3}{4}+\frac{1}{2}\epsilon_n\right)}
       {\Gamma\left(\frac{1}{4}+\frac{1}{2}\epsilon_n\right)}
  =-\tilde\kappa.
\end{equation}
Since the $\Gamma$-function has poles for negative integer values, one
can immediately see that the energy eigenvalues for a strong central
barrier $(\kappa\rightarrow\infty)$ converge to
$E_n\rightarrow\frac{3}{2}\hbar\omega, \frac{7}{2}\hbar\omega,
\frac{11}{2}\hbar\omega,\dots$. They become identical to the values of
the odd eigenfunctions in the limit $\kappa=\infty$, leading to a
double degeneracy of all eigenstates \cite{Wilkens:98}.
Fig.~\ref{fig:GSkappa} displays the square of the ground state
eigenfunctions of the Hamiltonian in eq.~\eqref{eq:deltaHamiltonian}
for different values of $\kappa$. One can clearly see the development
of a node at $x=0$ for increasing strength of the splitting potential.

In the limit $\kappa=\infty$, when the neighbouring even and odd
eigenfunctions become degenerate, the Whittaker functions of the even
states simplify to Hermite polynomials with an absolute value
argument.

\section{Quantum Gas in a $\delta$-split harmonics trap}
\label{sec:QuantumGas}

To find the exact many-particle wave-function for bosons or fermions,
one has to calculate the Slater determinant from the single particle
eigenfunctions $\phi_n(x)$ and, for the bosons, apply the
symmetrization procedure to it. Due to the complicated structure of
the Whittaker functions for finite values of $\kappa$ this can only be
done numerically. However, for the limiting case of an infinitively
high central $\delta$--function this can in fact be done analytically.
The eigenfunctions are in this case given by
\begin{subequations}
 \begin{align}
   \label{eq:Hodd}
   \psi_n(x)&=C_ne^{-\frac{x^2}{2}}H_n(x)
   &&\quad\text{for}\; n\; \text{odd},\\
   \psi_n(x)&=C_{n+1}e^{-\frac{|x|^2}{2}}H_{n+1}(|x|)
   &&\quad\text{for}\; n\; \text{even},
 \end{align}
\end{subequations}
with the normalization $C_n=({\sqrt{\pi}a_0 2^n n!})^{-\frac{1}{2}}$,
and the Slater determinant is defined by
\begin{equation}
  \label{eq:SlaterDet}
    \psi_{F}(x_1,\dots,x_N)=
    \frac{1}{\sqrt{N!}}\det_{(n,j)=(0,1)}^{(N-1,N)}\psi_n(x_j).
\end{equation}
Inserting the explicit expressions for the Hermite polynomials into
the determinant and applying basic multi-linear operations, the
determinant can be reduced to
\begin{align}
  \det_{(n,j)=(0,1)}^{(N-1,N)}&\psi_n(x_j)=
    C\;2^{\frac{N^2}{2}}
  \left[\prod_j e^{-\frac{x_j^2}{2}}\right]\times\nonumber\\
  \times&
  \begin{vmatrix}
    |x_1|&x_1&|x_1|^3&x_1^3
    &\dots&|x_1|^{n-1}&x_1^{n-1}
    \\
    \\\dots&\dots&\dots
     &\dots&\dots&\dots&\dots
    \\
    \\ 
    |x_n|&x_n&|x_n|^3&x_n^3
   &\dots&|x_n|^{n-1}&x_n^{n-1}
  \end{vmatrix}
\end{align}
where $C=\prod_{n=1}^{N/2} C_{2n-1}^2$.  Unfortunately in this form
the determinant cannot be easily solved, due to the appearance of the
absolute values.

We can, however, rewrite the problem in a more amenable form. It is
easy to see that for the case of $\kappa=\infty$ due to the double
degeneracy of each energy level, one can choose a new set of
eigenfunctions given by
$\phi_n^\pm=\frac{1}{\sqrt2}(\psi_{2n}\pm\psi_{2n+1})$
\cite{Wilkens:97}. These eigenfunction correspond to the particles
being physically located entirely either left or right of the barrier
and the wave-functions $\phi_n^\pm$ are mutually symmetric with
respect to $x=0$.  Therefore we can restrict our treatment to the
spatial area $x>0$ and consider only $\frac{N}{2}$ particles (for
simplicity we assume an even number of particles). The Slater
determinant can then be explicitly calculated by just employing the
odd Hermite polynomials of eq.~\eqref{eq:Hodd}
\begin{equation}
   \det_{(n,j)=(0,1)}^{(\frac{N}{2}-1,\frac{N}{2})}H_{2n+1}(x_j)
  = 2^{\frac{N^2}{8}}
    \left[\prod_j^{\frac{N}{2}} x_j\right]
    \prod_{(j,k)=(1,j+1)}^{(\frac{N}{2},\frac{N}{2})}(x_j^2-x_k^2)
\end{equation}
which immediately gives a simple and exact expression for the
fermionic many-particle wave-functions of eq.~\eqref{eq:SlaterDet}.
The bosonic form then follows directly from
eq.~\eqref{eq:Symmetrisation} and is given by
\begin{align}
   \label{eq:PsiB}
  \Psi_B(x_1,\dots,x_N)=\frac{\tilde C}{\sqrt{N!}}&  2^{\frac{N^2}{8}}
    \left[\prod_j^{N/2} e^{-\frac{x_j^2}{2}}|x_j|\right]\times\nonumber\\
    &\times\prod_{(j,k)=(1,j+1)}^{(N/2,N/2)}|x_j^2-x_k^2|
\end{align}
with $\tilde C=\prod_{n=1}^{N/2} C_{2n-1}$. It resembles the
wave-function found by Girardeau \textsl{et al.} \cite{Triscari:01}
for the undisturbed harmonic oscillator, and differs due to the reason
that we have constructed the Slater determinant only from the odd
Hermite polynomials.

\begin{figure}[tbp]
  \includegraphics[width=\linewidth]{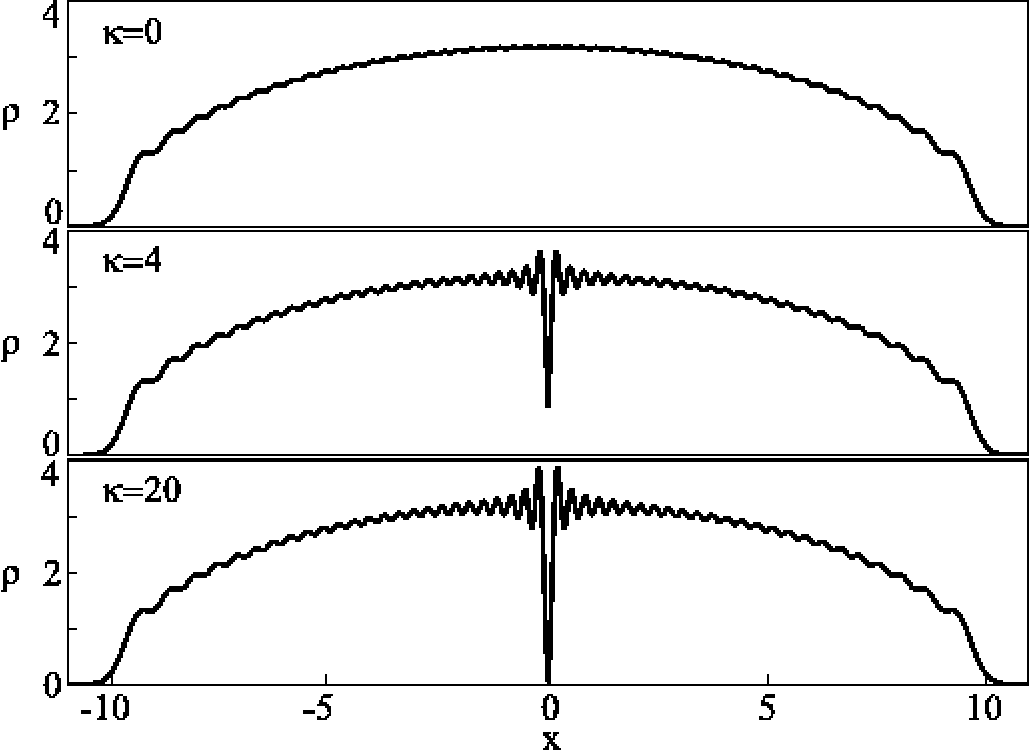}
  \caption{Ground state density for a one dimensional quantum gas of 50 particles in a
    $\delta$-split harmonic trap. The strength of the
    $\delta$-function is given by $\kappa$. Note the increase in
    density in the center of the trap around the notch. Note also that
    the appearance of the $\delta$-function has no influence on the
    size of the atomic cloud.}
 \label{fig:DensityN50}
\end{figure}

In Fig.~\ref{fig:DensityN50} the density of a cloud of $N=50$ bosons
is shown for different heights of the central potential. One can
clearly see that the central notch gets deeper with increasing values
of $\kappa$, but one also finds an increase in density of about 25\%
just around the hole for large values of $\kappa$. As one would
expect, the influence of the $\delta$-potential only extends a short
distance (the 'healing length') from $x=0$ and does not have any
influence on, say, the size of the atomic cloud.

\begin{figure}[bp]
  \includegraphics[width=\linewidth]{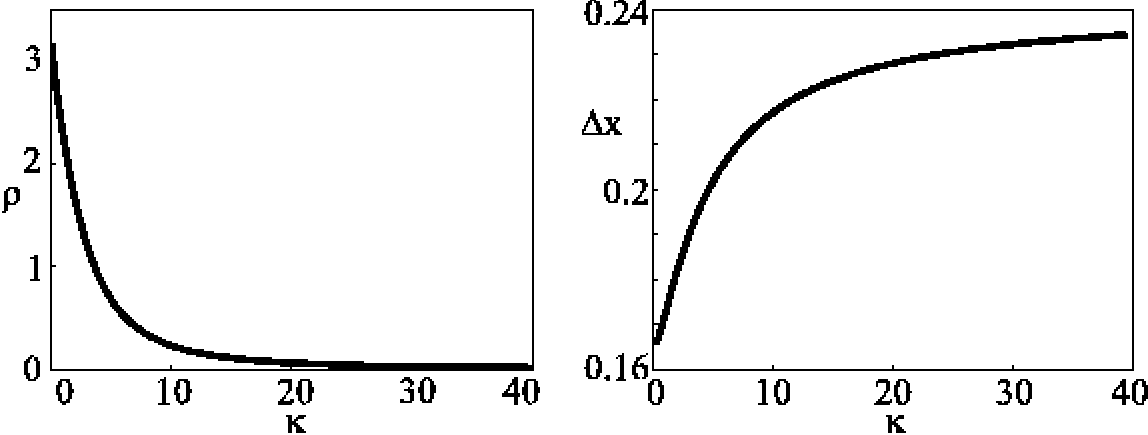}
  \caption{The left hand side shows the density of a gas at $x=0$ depending on 
    the strength of the delta function. The right hand side shows the
    FWHM of the central notch. Note that the width increase with
    increasing 'darkness'.}
 \label{fig:MinDen_FWHM}
\end{figure}

The density at $x=0$ of the gas as a function of $\kappa$ is displayed
in the LHS of Fig.~\ref{fig:MinDen_FWHM}. One can see that for values
of $\kappa$ of a few $\omega$ the density drops quickly towards zero.
On the RHS the full-width-half-max size of the central notch is shown.
As compared to the solitonic solutions for Bose gases described by the
Gross-Pitaevskii equation where the width of the notch decreases with
increasing darkness, the opposite behaviour is found here. This is of
course due to the fact that the density void of the soliton in a Bose
condensate is due to a phase jump in the wave-function and not an
external potential.

\section{Dark Solitons}
\label{sec:DarkSolitons}

As already mentioned in the last section, the density distributions
shown in Fig.~\ref{fig:DensityN50} for finite $\kappa$ resemble the
dark and grey soliton states of the Gross-Pitaevskii equation (GPE).
Dark (and grey) solitons are solutions to the GPE with repulsive
interaction and their existence in dilute Bose-condensates has been
experimentally demonstrated \cite{Lewenstein:99,Phillips:00}. Recently
also the appearance of dark solitons in TG gas has been examined in a
toroidal trap \cite{Wright:00-2}.

Having derived the exact many-particle state for the pierced harmonic
oscillator, we can now compare the TG 'notch' state with soliton
solutions of the Gross-Pitaevskii equation in a trap. Note, that we
have not only the 'dark notch' solution at hand, but we also know the
full family of solutions for 'grey notches', where the density is
reduced but does not go to zero.

We are, however, again not able to derive any analytical expression
for the grey soliton densities, since the summations over the
eigenstate densities involve a summation over $\epsilon_n$. This
doesn't represent a series of integers nor is there an easy
relationship between the integer $n$ and the $\epsilon_n$ and we
therefore resort to numerical simulations.

To simulate the time evolution of the bosonic gas, one can make use of
a time dependent version of the Fermi-Bose mapping theorem
\cite{Wright:00-2}. The many-particle fermionic wave-function is then
given by
\begin{equation}
  \psi_F(x_1,\dots,x_N;t)=C\det_{(i,j)=(0,1)}^{(N-1,N)}\phi_i(x_j,t),
\end{equation}
and the single particle eigenstates of the external potential evolve
under the time-dependent Schr\"odinger equation
\begin{equation}
  i\hbar\frac{\partial\phi_i(x,t)}{\partial t}=
  \left[-\frac{\hbar^2}{2m}\frac{\partial^2}{\partial x^2}+V(x,t)\right]\phi_i(x,t).
\end{equation}
Symmetrization of these states then leads to the appropriate solutions
for the bosonic case. To investigate the dynamics of the notch, we
assume that we can prepare the system in the stationary state we want
for fixed $\kappa$ and then we instantaneously switch $\kappa=0$.

\begin{figure}[tbp]
  \includegraphics[width=\linewidth]{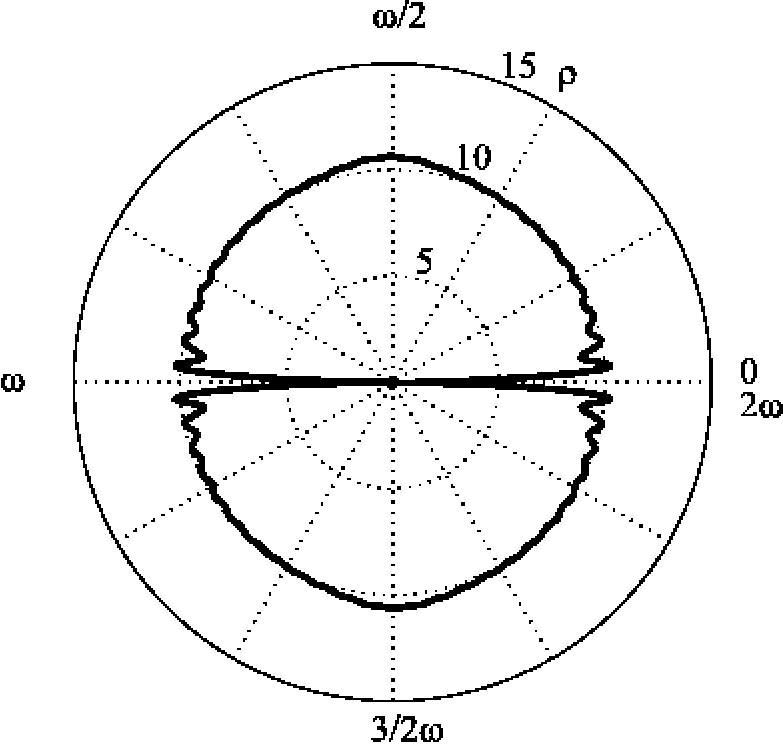}
  \caption{Density at $x=0$ during one oscillation period after $\kappa$ has
    instantaneously been changed from $\kappa=\infty$ to $\kappa=0$.
    One can clearly see the recurrence of the notch after one trap
    period.}
 \label{fig:Oscillation}
\end{figure}

First, it is easy to understand that the recurrence frequency for the
density notch in the gas is given by the trap frequency. Since the
density distribution is in particular the density distribution for
non-interacting fermions, after one trapping period $\omega$ every
fermion is back in the same state where it started and the notch is
recreated (see Fig.~\ref{fig:Oscillation}). Even though this result is
to be expected for non-interacting fermions, it is not at all obvious
for a system of strongly interacting bosons. In particular it is in
contrast to behaviour of dark solitons in the GPE, for which the
oscillation frequency is given by $\omega/\sqrt 2$ \cite{Anglin:00}.

Since the filling in and the oscillation of the dark notch as seen in
Fig.~\ref{fig:Oscillation} are purely due to the even eigenfunctions
(the odd ones are stationary states for any value of $\kappa$), one
could consider to stabilize the density notch by imprinting a
phase-step with $\Delta\Phi=\pi$ onto the sample. This would
artificially create the same stabilization mechanism that the soliton
solution of the GPE has, i.e.~it would transform the single particle
states with an even parity to states with an odd one.  However, since
one cannot make this process parity selective, at the same time it
would transfer the odd states into even states and nothing would be
won.

\section{Detection}
\label{sec:Detection}

\begin{figure}[tbp]
  \includegraphics[width=\linewidth]{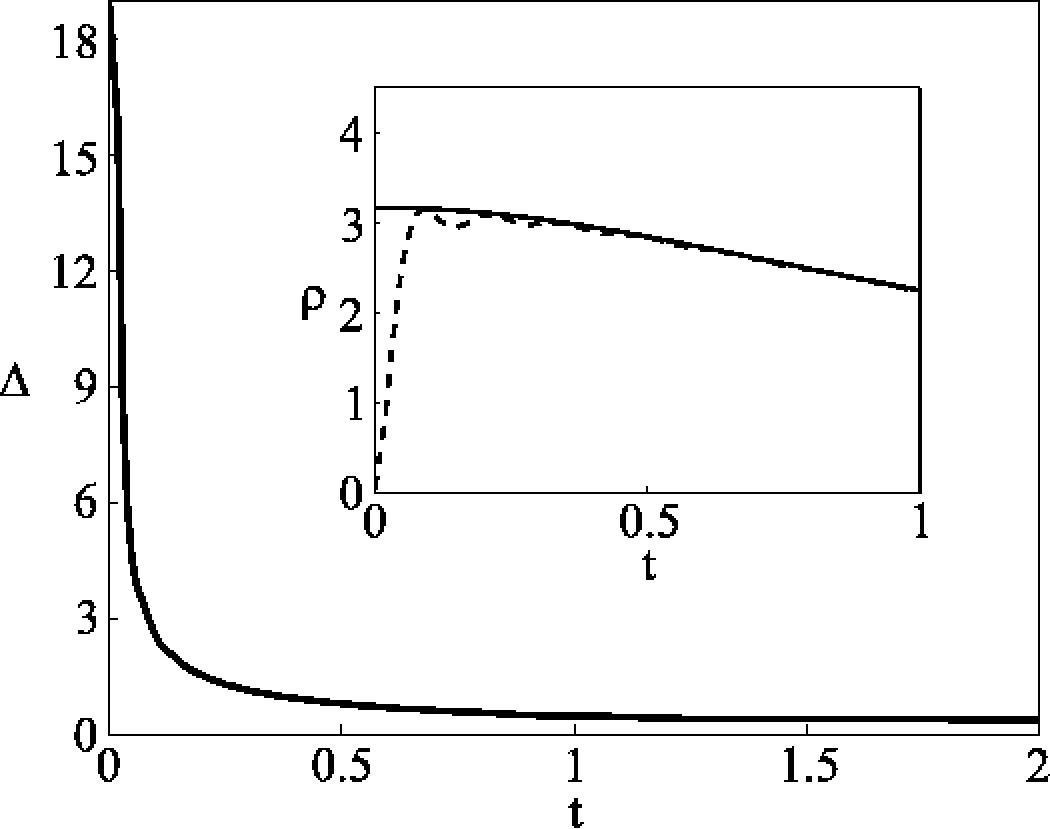}
  \caption{Integrated absolute difference between the density distributions of the 
    ground state gas for $\kappa=0$ and $\kappa=\infty$ during the
    free expansion. The fast decay is a sign of the
    indistinguishability of the two ground states via the density
    distributions.  The inset shows the density $\rho(x=0)$ for
    $\kappa=\infty$ (broken line) and for $\kappa=0$ (full line). One
    can see that the hole vanishes at very early stages during the
    free evolution. Due to the absence of an external potential, there
    are no recurrences here. The timescale in both graphs is in units
    of the inverse trap frequency. }
 \label{fig:FreeExpansion}
\end{figure}

The absence of a self-similar free expansion has been pointed out as
one criterion to distinguish the TG gas from a one-dimensional
Bose-gas in the Gross-Pitaevskii regime \cite{Santos:02}. Since the
atomic cloud is only disturbed on a very small length scale by the
central $\delta$-function, and, as can be seen from
Fig.~\ref{fig:Oscillation}, the lifetime for the density notch once
the central barrier is removed is much shorter than a trap frequency,
the modification due to the split trap has no measurable effect on the
size and dynamics of the expanding cloud (see
Fig.~\ref{fig:FreeExpansion}).  Especially, since the lifetime of the
density notch is so short, once cannot observe any broadening of it

This fact can also be seen by realizing that the energy increase for
finite $\kappa$ is at maximum $\Delta E\rightarrow
\frac{N}{2}\hbar\omega$ for the limiting case of
$\kappa\rightarrow\infty$, which is very small on the scale of the
total energy of the cloud, $E_{tot}=N^2/2$.  (Note that the chemical
potential alone is already $\mu=(N-1/2)\hbar\omega$.)

We suggest here a different approach that relies on the insensibility
of the state towards phase imprinting. In standard traps ($\kappa=0$)
phase imprinting has a substantial effect on the density distribution
of a sample, a fact which has been successfully utilized for creating
dark solitons in ground state Bose-Einstein condensates. In the
dynamics following the imprinting the gas tries to adjust its density
distribution to the phase, thereby exciting various phonon modes and
modifying the Thomas-Fermi density distribution. Recently phase
imprinting was theoretically investigated for fermionic samples, and
the emergence of solitonic structures was also found there
\cite{Rzazewski:02}.  However, since the phase evolution of the single
particles in a Fermi sea is not coherent, the effects found were less
pronounced.

To demonstrate the insensibility of the bosonic and fermionic states
in a $\delta$-split trap, we consider here the limiting case of
$\kappa\rightarrow\infty$, and the reaction to an imprinted phase of
$\Delta\phi=\pi$. Applying this imprinting only spatially selectively
for, say, $x<0$, one can see immediately that the eigenfunctions
$\psi_n(x)$ with odd values of $n$ turn into $\psi_{n-1}(|x|)$ and
vice versa. Since after the imprinting we again deal only with
stationary eigenfunctions, the phase imprinting (assuming an ideal
imprinting mask) does not have any observable effect on the density
distribution.  Mathematically one can see immediately that the Slater
determinant due to its multilinearity is unchanged by the imprinting
of the phase (for an even number of particles). Therefore the absence
of distortions could be used a a possible tool to detect the split
trap state.

\section{Experimental realization of the $\delta$-potential}
\label{sec:Realization}

Let us finally remark on the experimental realization of such a
potential, since the $\delta$-function is usually regarded as a
mathematical idealization far from experimental reality. A very
successful method to manipulate the density profile of an atomic cloud
is to employ dipole forces with appropriately detuned lasers.  It is
nevertheless experimentally quite challenging to focus and stabilize a
laser beam to such a small diameter that it represents a point-like
potential to the atoms as necessary for the situation considered here.
However, since low densities are best reached in shallow traps, the
ratio of the size of the single particle states vs.~the fwhm of the
focused laser beam increases with increasing trap size, thereby
creating a more and more localized potential. The tuning of $\kappa$
could in this case be done 'easily' by adjusting the intensity of the
laser beam.

Here we suggest an alternative method which at the same time also
allows for easy tuning of $\kappa$, even though it is also
experimentally more challenging.

If optical potentials do not suffice in terms of localization, the
obvious next step is to employ potentials made from matter waves.
Multicomponent condensates have been investigated in the last years
and the conditions for such systems to exhibit phase separation have
been determined. Depending on the internal energy of the atoms and on
their respective masses shell like structures have been predicted,
where one component occupies the inner part of the trap and the other
component occupies the outer part. By realizing a situation like this
and then reducing the number of atoms occupying the central part of
the trap, one can, in the ideal case, come to the situation where a
single atom is trapped in the trap center and the second component is
arranged around it.  This would correspond to an almost perfect
point-like potential for the atoms of the outer component.  With the
help of a selective Feshbach resonance the strength of the interaction
between the single atom and the gas could then be adjusted.

\section{Conclusions}
\label{sec:Conclusions}

We have investigated the behaviour of low density, one-dimensional
Bose and Fermi gases in a $\delta$-split harmonic trap. We have
derived the exact many-particle wave function for non-interacting
fermions and were able, due to the Fermi-Bose mapping theorem, to
determine also the exact expression for the many-particle
wave-function for the bosonic TG gas in the limiting case of an
infinitely high $\delta$-barrier. Both many-particle wave-functions
turned out to be simple products over quadratic terms. We have thereby
presented another of the rare examples of exactly solvable models for
many-particle systems.

We have further numerically investigated the dynamical behavior of
many-particle, split-trap eigenstate when the central barrier is
suddenly removed and compared it to the dynamics of the dark soliton
solutions in the GPE. The recurrence time for the 'notch' in the
linear low density limit was found to be the trapping frequency, which
is different to the oscillation frequency of the dark Gross-Pitaevskii
soliton of $\omega/\sqrt{2}$.

Finally, we have suggested using the systems insensitivity to phase
imprinting as a tool for detecting the presence of the notch-state and
employing a second component and a selective Feshbach resonance to
create and manipulate the point-like central potential.

\acknowledgements This work was financially supported by the European
Union under contract3 HPRN-CT-2000-00034-VISTA, by the Science
Foundation Ireland under the grant sfi/01/fi/co and by the Royal Irish
Academy.  Discussions with S.~Nic Chormaic and J.~Denschlag are
gratefully acknowledged. TB would like to thank John McInerney for his
support.

\end{document}